\documentclass[a4paper,11pt]{article}
\usepackage{pos}
\usepackage{subfigure}
\usepackage{wrapfig}
\NewDocumentCommand\Nf{mgg}{N\textsubscript{f}=#1\IfNoValueTF{#2}{}{+#2}\IfNoValueTF{#3}{}{+#3}}
\NewDocumentCommand\vol{mg}{#1\textsuperscript{3}\IfNoValueTF{#2}{}{×#2}}

\newcommand{\tins}{t_\mathrm{ins}}
\newcommand{\xins}{x_\mathrm{ins}}
\newcommand{\vxins}{\vec{x}_\mathrm{ins}}

\makeatletter
\let\oldthebibliography\thebibliography
\renewcommand{\thebibliography}[1]{
  \oldthebibliography{#1}
  \setlength{\itemsep}{6pt}
}
\makeatother

\title{Nucleon strange electromagnetic form factors using \Nf{2}{1}{1}
 twisted-mass fermions at the physical point}
\ShortTitle{Nucleon strange EM form factors}
\usepackage{makecell}

\author[a,b]{Constantia Alexandrou}
\author[b]{Simone Bacchio}
\author[c]{Mathis Bode}
\author[d]{Jacob Finkenrath}
\author[d]{Andreas Herten}
\author[a,b]{Christos Iona}
\author[b]{Giannis Koutsou}
\author[b]{Ferenc Pittler}
\author*[b]{Bhavna Prasad}
\author[a]{Gregoris Spanoudes}

\affiliation[a]{Department of Physics, University of Cyprus}
\affiliation[b]{Computation-based Science and Technology Research Center, The Cyprus Institute}
\affiliation[c]{J\"{u}lich Supercomputing Centre, Forschungzentrum J\"{u}lich}
\affiliation[d]{Department of Physics, Bergische Universit\"{a}t Wuppertal}

\abstract{We present the strange electromagnetic form factors of the nucleon using lattice QCD with \Nf{2}{1}{1} twisted mass clover-improved fermions and quark masses tuned to their physical values. Using four ensembles with lattice spacings of $a=0.080$ fm, $0.068$ fm, $0.057$ fm and $0.049$ fm, and similar physical volume, we obtain the continuum limit directly at the physical pion mass. The disconnected strange contributions are computed using high statistics two-point functions combined with stochastic noise mitigation techniques, such as spin-color dilution and hierarchical probing in the estimation of the quark loop. From the momentum dependence of the form factors, we provide the strange electric and magnetic radii, as well as the strange magnetic moment in the continuum limit.}

\FullConference{The 42nd International Symposium on Lattice Field Theory (LATTICE2025)\\
2-8 November 2025\\
Tata Institute of Fundamental Research, Mumbai, India\\}

\begin{document}

\maketitle

\section{Introduction}

The strange contribution to the electromagnetic form factors of the nucleon gives us a direct probe to the underlying sea quark dynamics in the nonperturbative regime of quantum chromodynamics (QCD) as it constitutes the lightest non-valence quark in the nucleon. Experimentally, the strange electromagnetic form factors are indirectly measured using parity-violating asymmetry measurements which originate from the interference of electromagnetic and weak interactions due to electroweak mixing~\cite{Kaplan:1988ku, Maas:2017snj}. A considerable experimental effort has been undertaken over the last two decades to measure the strange contribution to the electromagnetic form factors of the nucleon via the asymmetry cross-section measurements, such as SAMPLE~\cite{SAMPLE:2003wwa, Beise:2004py}, A4~\cite{A4:2004gdl,Maas:2004dh,Baunack:2009gy}, HAPPEX~\cite{HAPPEX:2005qax,HAPPEX:2005zgj,HAPPEX:2006oqy,HAPPEX:2011xlw}, and G0~\cite{G0:2005chy,G0:2009wvv}. Experimental data so far cannot  exclude a zero value for the strange magnetic moment and the electric and magnetic radii. Lattice QCD provides a unique opportunity to nonperturbatively compute the form factors very precisely and has thus been used as an additional constraint in experimental determinations of the weak charge of the proton, $Q_{\rm weak}$~\cite{Qweak:2018tjf}. In the last decade, with advancements aimed at mitigating the noise of the disconnected three-point functions, several lattice QCD results have been presented~\cite{Djukanovic:2019jtp,Alexandrou:2019olr,Sufian:2016pex,Green:2015wqa}. However, the results either have not been extrapolated to the continuum or used several heavier-than-physical pion mass ensembles  to reach the chiral and continuum limit. 
In these proceedings, we provide continuum limit results (see Ref.~\cite{Alexandrou:2026ait,Alexandrou:2026xcl} for more details) for the strange electromagnetic form factors of the nucleon using  four ensembles of clover-improved twisted mass fermions with two degenerate light,
strange, and charm quarks (\Nf{2}{1}{1}) with masses tuned to their
physical values, referred to as \textit{physical point} ensembles.

\section{Nucleon Electromagnetic form factors}
\label{sec:theory}
The electromagnetic form factors can be extracted from the ground state of the nucleon matrix element with the electromagnetic current insertion, where we consider the SU(2) flavor isospin symmetric limit, namely that the u–d quark mass difference and isospin-breaking effects arising from QED interactions are neglected. In Minkowski space the matrix element is given by,
\begin{gather}
 \langle N(p',s') \vert j_\mu \vert N(p,s) \rangle = \sqrt{\frac{m_N^2}{E_N(\vec{p}\,') E_N(\vec{p})}}
\bar{u}_N(p',s') \left[ \gamma_\mu F_1(q^2) + \frac{i \sigma_{\mu\nu} q^\nu}{2 m_N} F_2(q^2) \right] u_N(p,s)\, 
 \label{eq:me}
\end{gather}
where $N(p,s)$ is the nucleon with initial (final) momentum $p$ ($p'$)
and spin $s$ ($s'$), with energy $E_N(\vec{p})$ ($E_N(\vec{p}\,') $)
and mass $m_N$, $u_N$ is the nucleon spinor, 
$j_\mu$ is the vector current, and $q^2{\equiv}q_\mu q^\mu$ is the momentum transfer
squared with $q_\mu{=}(p_\mu'-p_\mu)$. The local vector current for the strange contribution to the nucleon is given by, 
$ j_\mu = e_s j^s_\mu = e_s \bar{\psi}_s \gamma_\mu \psi_s$
where $s$ denotes the strange quark flavor and $e_s=-1/3$ is the strange electric charge, which has not been multiplied into the final results. 
The  Dirac ($F^s_1$) and Pauli ($F^s_2$) form factors can be recombined in terms of the electric and magnetic Sachs form factors $G^s_E(q^2)$ and $G^s_M(q^2)$ as follows
\begin{gather}
  G^s_E(q^2) = F^s_1(q^2) + \frac{q^2}{4m_N^2} F^s_2(q^2),\,\,\,
  G^s_M(q^2) = F^s_1(q^2) + F^s_2(q^2)\,.\label{eq:sachs}
\end{gather}
The electric and magnetic
mean-squared radii are defined as the slope of the
corresponding Sachs form factor as $q^2\rightarrow 0$, namely
$ \langle r_X^2 \rangle^s = -6
  \frac{\partial G_X^s(q^2)}{\partial q^2} \Big
  \vert_{q^2=0},\,$
with $X=E,M$.

\section{Lattice setup}
\begin{figure}
\centering
  \includegraphics[width=0.3\textwidth]{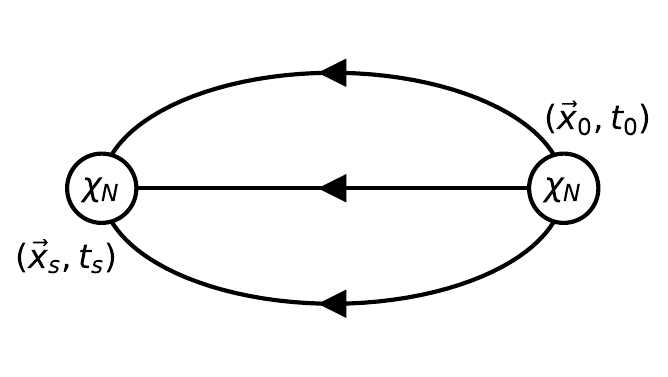}
  \includegraphics[width=0.3\textwidth]{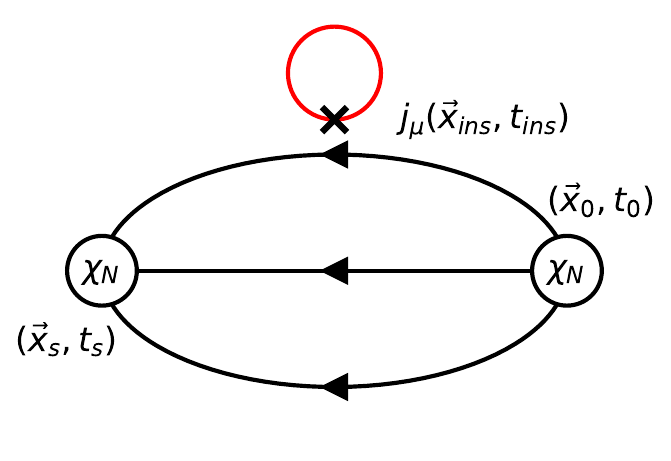}
  \caption{Nucleon two-point function (left) and disconnected nucleon three-point function (right).}
  \label{fig:diagrams}
\end{figure}

In order to access the desired nucleon matrix element in lattice QCD, we compute the two- and three-point correlation
functions as shown in Fig.~\ref{fig:diagrams}. In  Euclidean space-time with $Q^2=-q^2$, the two-point function is given by
\begin{gather}
C(\Gamma_0,\vec{p};t_s,t_0) = \sum_{\vec{x}_s} 
e^{{-}i (\vec{x}_s{-}\vec{x}_0) \cdot \vec{p}}
\mathrm{Tr} \left[ \Gamma_0 {\langle}\chi_N(t_s,\vec{x}_s) \bar{\chi}_N(t_0,\vec{x}_0) {\rangle} \right] \,, \label{eq:twop}
\end{gather}
where $\chi_N(\vec{x},t)$ is the standard nucleon interpolating field,
$\chi_N(\vec{x},t)=\epsilon^{abc}u^a(x)[u^{b\intercal}(x)\mathcal{C}\gamma_5d^c(x)]\,,$ and 
 $\mathcal{C}{=}\gamma_0 \gamma_2$ is the charge conjugation
matrix. The three-point function is given by
\begin{gather}
  C_\mu(\Gamma_\nu,\vec{q},\vec{p}\,';t_s,\tins,t_0)= 
  \sum_{\vxins,\vec{x}_s}  e^{i (\vxins {-} \vec{x}_0)  \cdot \vec{q}}  e^{-i(\vec{x}_s {-} \vec{x}_0)\cdot \vec{p}\,'} 
  \mathrm{Tr} [ \Gamma_\nu \langle \chi_N(t_s,\vec{x}_s)  j_\mu(\tins,\vxins) \bar{\chi}_N(t_0,\vec{x}_0) \rangle].
  \label{eq:thrp}
\end{gather}
The initial lattice site where the nucleon is created is denoted by  $x_0$ and referred to as the
\textit{source}, the lattice site where the current couples to a quark  $\xins$ as the \textit{insertion}, and the site where the nucleon is annihilated  $x_s$ as
the \textit{sink}. $\Gamma_\nu$ is a projector acting on Dirac
indices, with $\Gamma_0 {=} \frac{1}{2}(1{+}\gamma_0)$ yielding the
unpolarized and $\Gamma_k{=}\Gamma_0 i \gamma_5 \gamma_k$ the
polarized matrix elements. Without loss of generality we will take
$t_s$ and $\tins$ relative to the source time $t_0$ in what follows.

In order to extract the ground-state nucleon matrix element and cancel overlaps of the interpolating operator with the nucleon states, an
optimized ratio composed of two- and three-point functions is used, given by
\begin{align}
R_{\mu}(\Gamma_{\nu},\vec{p},\vec{p}\,';t_s,\tins) = \frac{C_{\mu}(\Gamma_{\nu},\vec{p},\vec{p}\,';t_s,\tins\
)}{C(\Gamma_0,\vec{p}\,';t_s)} 
\sqrt{\frac{C(\Gamma_0,\vec{p};t_s-\tins) C(\Gamma_0,\vec{p}\,';\tins) C(\Gamma_0,\vec{p}\,';t_s)}{C(\Gamma_0,\vec{p}\,';t_s-\tins) C(\Gamma_0,\vec{p};\tins) C(\Gamma_0,\vec{p};t_s)}}.
\label{eq:full_ratio}
\end{align}
In the limit of large time separations $\Delta E(t_s-\tins) \gg 1$ and $\Delta E\tins
\gg1$, the ratio in Eq.~(\ref{eq:full_ratio}) converges to the nucleon
ground state matrix element, where $\Delta E$ is the energy gap between the first excited state and the ground state.

\begin{table}[ht]
    \caption{Parameters of the four \Nf{2}{1}{1} ensembles analyzed
      in this work. From the leftmost to rightmost columns, we provide
      the name of the ensemble and its short acronym in parenthesis, the lattice volume, $\beta=6/g^2$ with
      $g$ the bare coupling constant, the lattice spacing, the pion
      mass, the value of $m_\pi L$, the number of configurations ($n_{\rm conf}$), and the number of source positions used ($n_{\rm src}$). Lattice spacings and pion
      masses are taken from
      Refs.~\cite{ExtendedTwistedMass:2022jpw,ExtendedTwistedMass:2024nyi}.}
    \label{tab:ens}
    \centering
    \begin{tabular}{cccccccc}
    \hline\hline
       Ensemble   & $(\frac{L}{a})^3{\times}(\frac{T}{a})$ & $\beta$ & \makecell[c]{$a$\\$[$fm$]$} & \makecell[c]{$m_\pi$\\ $[$MeV$]$}  & $m_\pi L$ & $n_{\rm conf}$ & $n_{\rm src}$\\
       \hline 
        \texttt{cB211.072.64} (\texttt{B})& $64^3 {\times} 128$ & 1.778 &  0.07957(13) & 140.2(2) & 3.62 & 749 & 349 \\
        \texttt{cC211.060.80} (\texttt{C})& $80^3 {\times} 160$ & 1.836 &  0.06821(13) & 136.7(2) & 3.78 & 399 & 650 \\
        \texttt{cD211.054.96} (\texttt{D})& $96^3 {\times} 192$ & 1.900 &  0.05692(12) & 140.8(2) & 3.90 & 493 & 368 \\
        \texttt{cE211.044.112} (\texttt{E})& $112^3 {\times} 224$ & 1.960 &  0.04892(11) & 136.5(2) & 3.79 & 501 & 339 \\
        \hline
    \end{tabular}
\end{table}

We use ensembles simulated with \Nf{2}{1}{1} twisted mass,
clover-improved fermions with quark masses tuned to approximately
their physical values. A summary of the parameters for the ensembles
is provided in Table~\ref{tab:ens}. We employ Gaussian smearing with APE-smeared gauge links to increase the overlap of the nucleon interpolating field with the nucleon ground state. For the strange matrix elements considered here, the quark field contractions of Eq.~(\ref{eq:thrp}) give rise to the disconnected three-point function shown in Fig.~\ref{fig:diagrams}. The strange quark loop is given by,
\begin{gather}
L_s(t_\mathrm{ins},\vec{q})=
\sum_{\vec{x}_\mathrm{ins}}e^{i\vec{q}\cdot\vec{x}_\mathrm{ins}}\mathrm{Tr}[D^{-1}_s(x_\mathrm{ins};x_\mathrm{ins})\gamma_\mu],
\label{eq:gen-one-end}
\end{gather}
which we compute using full dilution in spin and color and
hierarchical probing using a 4-dimensional coloring of distance
eight. Additionally, the \textit{generalized one-end trick} arising
from the properties of the twisted mass fermions is also used. The
quark loop can then be evaluated for every time slice $\tins$. For
computing the disconnected three-point function, we correlate the
quark loops with two point functions for all $t_s$ and $\tins$, as
well as finite values of sink momenta ($p'$), in addition to
$p'$=0. Details on the number of source positions ($n_{\rm src}$) used
in the computation of the two- and the three-point functions as well
as the number of configurations ($n_{\rm conf}$) available per
ensemble are also tabulated in Table~\ref{tab:ens}.

The strange disconnected three-point functions are computed using the
local vector current and therefore need to be renormalized. We refer
to Ref.~\cite{Alexandrou:2025vto} for details on the approach for
computing the renormalization constant of the vector current. In the
last column of Table~\ref{tab:disc_range}, the renormalization
constants used in this work for all ensembles are tabulated.

\section{Extraction of form factors}

\begin{figure}
    \centering
    \includegraphics[width=0.49\textwidth]{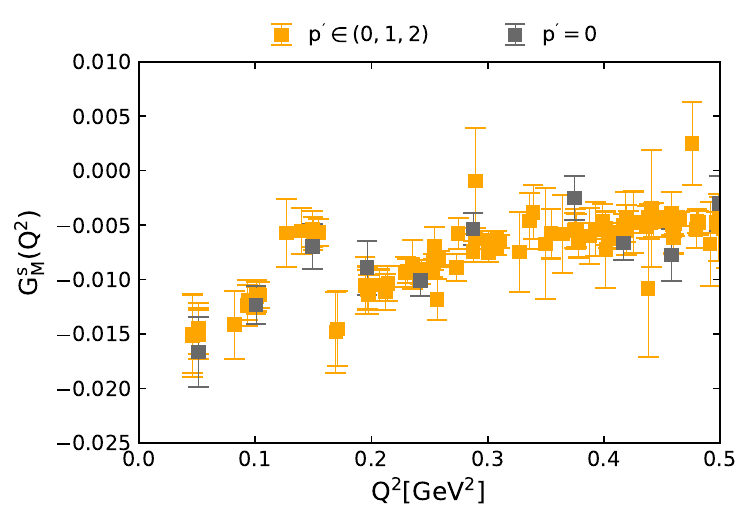}
    \includegraphics[width=0.49\textwidth]{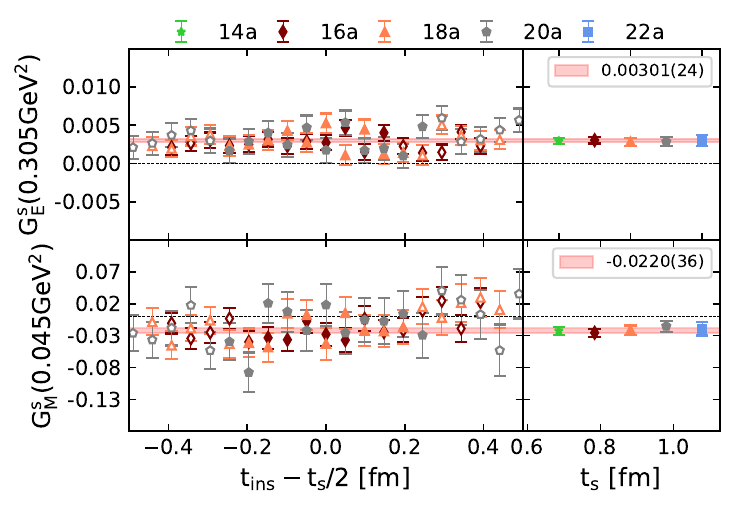}
    \caption{Left: $G^s_M(Q^2)$ for the ensemble
      \texttt{cC211.60.80} obtained by employing the lab and boosted frames with sink momenta
      $p'^2\in(0,1,2)$ (orange squares) and by only using the lab frame 
      (gray squares). Right: The renormalized $G_E^s(Q^2=0.305 \rm \;GeV^2)$ (top) and $G_M^s(Q^2=0.045 \rm \;GeV^2)$ (bottom) for the \texttt{cE211.044.112} ensemble. The left column shows source-sink separations $t_s = \rm 16a, 18a$ and $20\rm a$ as indicated in the header, for all insertion times $\tins$, where the filled points are incorporated in the plateau fits. The right column shows the result using plateau fits for $t_s$ as indicated in the header. The red band denotes the final value.}
    \label{fig:pp0_pp2_compare_E_disc}
\end{figure}

Since additional sink momenta are available to us at no additional
cost, we use $\vec{p}'=\frac{2\pi}{L}\vec{k}$ for $\vec{k}^2=0$, 1,
and 2 in a similar way to our approach for the disconnected light
contributions in Ref.~\cite{Alexandrou:2025vto}. In order to extract
the bare form factors for each value of the momentum transfer squared
($Q^2$), one needs to solve a set of equations arising from different
combinations of $\Gamma_\nu$ and $\mu$ depending on the momenta
$\vec{p}'$ and $\vec{q}$ entering the optimized ratio,
$R_\mu(\Gamma_\nu;\vec{p}', \vec{q})$ of
Eq.~(\ref{eq:full_ratio}). The addition of sink momenta renders the
set of equations inseparable for the two form factors necessitating
the use of a Singular Value Decomposition (SVD) to solve them (see
Appendix of Ref.~\cite{Alexandrou:2026ait}). In the left plot of
Fig.~\ref{fig:pp0_pp2_compare_E_disc}, we demonstrate the advantage of
having additional sink momenta by showing a comparison of the increase
in the number of $Q^2$ values with the additional sink momenta when
compared to $\vec{p}'=0$ for the \texttt{cC211.60.80} ensemble.

To obtain the time-independent ground-state contribution to the ratio,
i.e.  $R^\mu(\Gamma_\nu;\vec{p}', \vec{q}, \tins,
t_s)\allowbreak\xrightarrow[t_s-\tins\gg]{t_s\gg} \Pi^\mu(\Gamma_\nu;\vec{p}',
\vec{q})$, we perform \textit{Plateau fits} to the optimized ratio in
Eq.~(\ref{eq:full_ratio}). In our analysis, we observe that the
results from the plateau fits exhibit mild to no variation with change
in the source-sink separation, as shown in the right plot of
Fig.~\ref{fig:pp0_pp2_compare_E_disc}. We hence take a weighted average of results from plateau fits as our final result.

\begin{wraptable}[10]{r}{0.6\textwidth}
    \vspace{-1em}
    \centering
    \caption{Three-point
      function source-sink separation ($t_s$), the minimum insertion time ($\tins^{\rm min}$) with the maximum insertion time being $\tins^{\rm max}=t_s-\tins^{\rm min}$ used to obtain results for each ensemble. We also provide the renormalization constants ($Z_V$) used.} 
    \label{tab:disc_range}
    \begin{tabular}{cccc}
    \hline\hline Ensemble & $t_s/a$ & $\tins^{\rm min}/a$ & $Z_V$ \\
    \hline 
     B & 10, 12, 14, 16  & 3 & 0.7228(5)\\
     C & 12, 14, 16, 18  & 3 & 0.7373(5)\\
     D & 14, 16, 18, 20, 22 & 4 &  0.7521(4)\\
     E & 16, 18, 20, 22, 24, 26 & 5 & 0.7638(4)\\
    \hline 
    \end{tabular}
\end{wraptable}
In our fitting procedure, we first fit
the two-point function at $\vec{p}^2=0$ using a three-state fit to
extract the ground-state energy, $E_0(0)$, starting with similar
physical source-sink seperations, across all ensembles which is then
taken as a constant value and used in the evaluation and
pseudo-inversion of the coefficient matrix using SVD per jackknife
bin. A similar analysis is carried out for all the ensembles and the
fit ranges used to obtain the form factors at all $Q^2$ values has
been provided in Table~\ref{tab:disc_range}.

\section{Results for form factors}

\begin{figure}
    \centering
    \includegraphics[width=\textwidth]{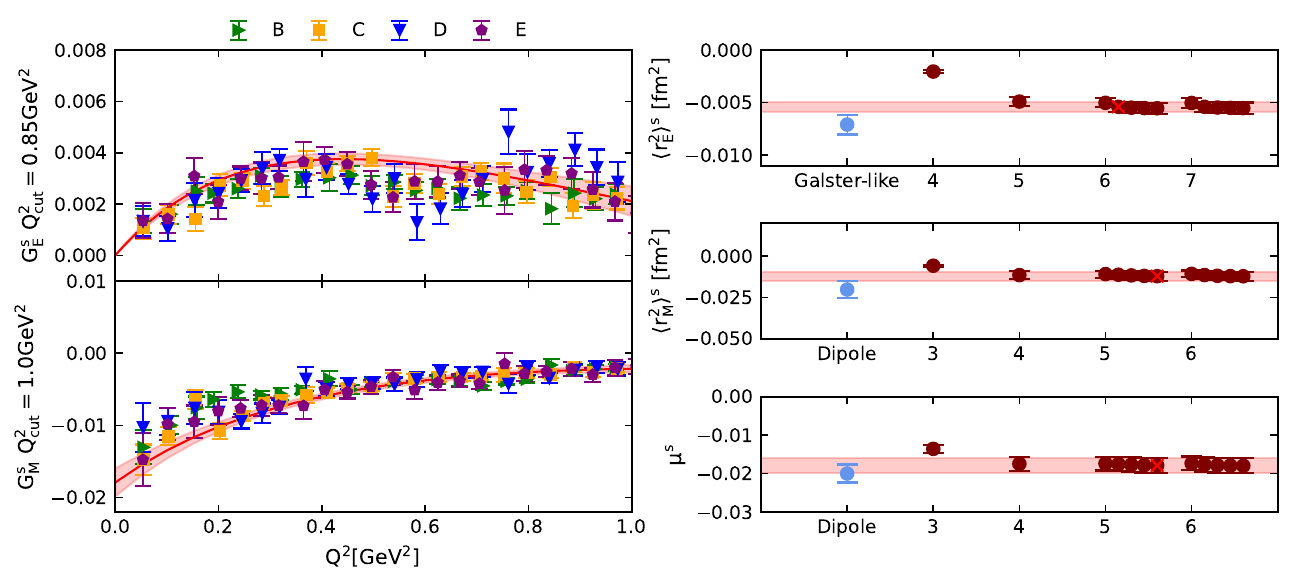}
    \caption{$G_E^s(Q^2)$ (top left) and $G_M^s(Q^2)$ (bottom left) for all ensembles indicated in the header. The red band corresponds to the continuum limit obtained by \textit{one-step} $z$-expansion fit with $Q^2_{\rm cut}=0.85\rm \;GeV^2$ for $G_E(Q^2)$ and $Q^2_{\rm cut}=1.0\rm \;GeV^2$ for $G_M(Q^2)$. In the right panel, the values of strange electric (top) and magnetic radii (middle) and the magnetic moment (bottom) as a function of the dependence of the the $z$-expansion order ($k_{\rm max}$) and prior widths
	($w$) (bottom panel). The red cross and the red band running in the three right panels corresponds to the final result chosen for each observable. The blue point is the result from the one-step dipole (Galster-like) fit for the magnetic (electric) form factor, shown for comparison.}
    \label{fig:GE_GM_dipole}
\end{figure}

In order to obtain the strange electric and magnetic radii, and the magnetic moment, we need to parameterize the $Q^2$ dependence of the form factors. Similar to Ref.~\cite{Alexandrou:2025vto}, we augment the parameters of the standard functional form to incorporate the cut-off effects by including the $a^2$-dependence, thereby obtaining results directly at the continuum limit, i.e., the \textit{one-step} approach. We additionally fit each ensemble seperately to the unchanged parameterizations and obtain the continuum limit by taking a linear extrapolation in $a^2$ called the \textit{two-step} approach. We use a dipole ansatz for the magnetic form factor,
\begin{gather}
    G(Q^2) = \frac{g}{\left[1+\frac{Q^2\langle r'^2 \rangle}{12}\right]^2}
    \xrightarrow[\langle r'^2(a^2)\rangle=\langle r'^2\rangle_0+a^2\langle r'^2\rangle_2]{g(a^2)=g_0+a^2g_2}
    G(Q^2,a^2) = \frac{g(a^2)}{\left[1+\frac{Q^2}{12}\langle r'^2(a^2)\rangle \right]^2},
    \label{eq:dipole}
\end{gather}
a Galster-like parameterization for the electric,
\begin{gather}
  G(Q^2) = \frac{Q^2 A}{4 m_N^2 + Q^2 B}\frac{1}{\left(1+\frac{Q^2}{0.71 {\rm GeV}^2}\right)^{2}}
  \xrightarrow[B(a^2)=B_0+a^2B_2]{A(a^2)=A_0 + a^2A_2}
  G(Q^2, a^2) = \frac{Q^2 A(a^2)}{4 m_N^2 + Q^2 B(a^2)}\frac{1}{\left(1+\frac{Q^2}{0.71 {\rm GeV}^2}\right)^{2}}
  \label{eq:Galster-like}
\end{gather}
and a $z$-expansion parameterization for both electric and magnetic,
\begin{gather}
G(Q^2) = \sum_{k=0}^{k_{\rm max}} c_k z^k(Q^2)
\xrightarrow[]{c_k(a^2)=c_{k,0}+a^2c_{k,2}}
G(Q^2,a^2) = \sum_{k=0}^{k_{\rm max}} c_k(a^2) z^k(Q^2),
\label{eq:zexp}
\end{gather}
\begin{equation}
\text{where, }\;\;\; z(Q^2) = \frac{\sqrt{t_{\rm cut} + Q^2} - \sqrt{t_{\rm cut}+t_0} }{
  \sqrt{t_{\rm cut} + Q^2} + \sqrt{t_{\rm cut}+t_0} }
\label{eq:zQ2}
\end{equation}
with $t_{\rm cut}=(2m_K)^2$ being the particle production threshold,
where, the kaon mass, $m_K=486 \rm \;MeV$ and $t_0=0$.  Additionally,
we require smooth convergence of the form factor to zero at
$Q^2\rightarrow \infty$ and employ gaussian prior with a width, $w$,
which we vary together with the $k_{\max}$.

\begin{figure}
  \centering
    \includegraphics[width=0.475\textwidth]{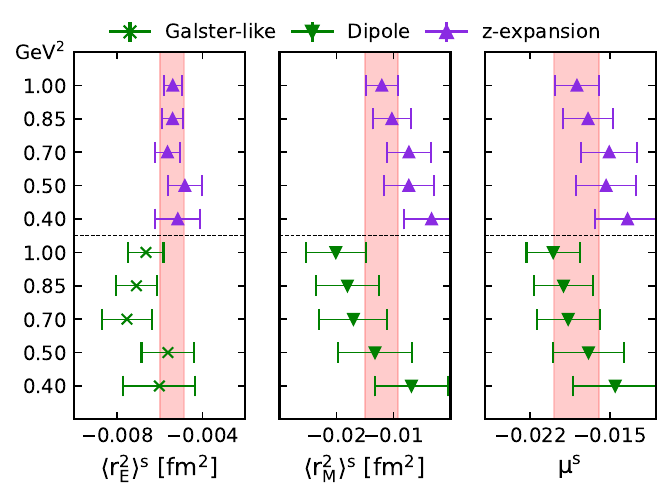}
    \includegraphics[width=0.51\textwidth]{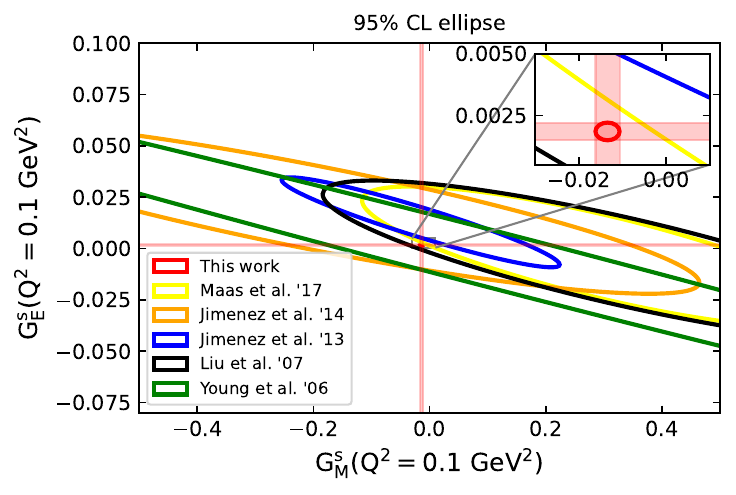}
    \caption{Left: Values of $\langle r^2_E\rangle^s$, $\langle r^2_M\rangle^s$ and $\mu^s$ obtained as a result of fitting to $G_E^s$ and $G_M^s$ with different $Q^2$ parameterizations and $Q^2_{\rm cut}$. The upward-pointing violet triangles denotes results from the $z$-expansion fits, green crosses denotes the Galster-like fit and downward-pointing green triangles denotes the results from dipole fits. The red band that runs through vertically corresponds to the model-averaged value using AIC. Right: Ellipses showing $95\%$ confidence curves from review of different experimental data with green ellipse from Ref.~\cite{Young:2006jc}, the orange from Ref.~\cite{Gonzalez-Jimenez:2014bia}, the black from Ref.~\cite{Liu:2007yi} and the blue from Ref.~\cite{Gonzalez-Jimenez:2011qkf}.
      The red bands and the red ellipse in the inset indicates the values extracted in this work.  }
    \label{fig:Results}
\end{figure}

\begin{wrapfigure}[15]{R}{0.5\textwidth}
    \vspace{-1em}
    \centering
    \includegraphics[width=\linewidth]{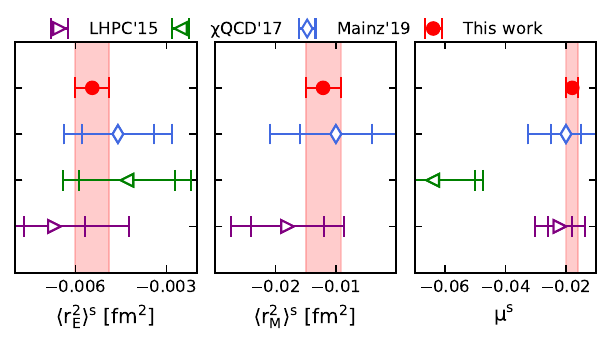}
    \caption{Results for the strange electric and magnetic radii and the
      magnetic moments of the nucleon obtained within this
      work (red circles). The light vertical bands correspond to
      statistical error. We compare to previous results by LHPC (right-pointing open triange)~\cite{Green:2015wqa}, $\chi$QCD (left-pointing open triangle)~\cite{Sufian:2016pex} and the Mainz group (blue diamond)~\cite{Djukanovic:2019jtp}.}
    \label{fig:Compare_lattice_results}
\end{wrapfigure}
For the Galster-like and dipole ansatz we use both the one- and two-step approaches to confirm the continuum limit values of the radii and magnetic moment are compatible between the two. We also see a mild linear dependence in $a^2$ reinforcing the need to include a lattice spacing dependence. We proceed to use the \textit{one-step} approach as it defines one global reduced $\chi^2$ value, providing direct input to model averaging procedure. In Fig.~\ref{fig:GE_GM_dipole}, we show our results on the electric and magnetic form factors for the four ensembles as well as fits using the $z$-expansion in the continuum limit. We also show results on the strange electric ($\langle r^2_E\rangle^s$) and magnetic ($\langle r^2_M\rangle^s$) radii and magnetic moment ($\mu^s$) and present a comparison for results obtained by varying the gaussian prior width $w$ and the order of the $z$-expansion, $k_{\rm max}$. The blue point corresponds to the results obtained from one-step dipole and Galster-like forms, presented only for comparison. The point where we observe convergence, indicated by the red cross and red horizontal bands, are taken as the final values for the three quantities. We additionally vary the mometa cuts $Q^2_{\rm cut}$, and model average using Akaike Information Criterion (AIC), as shown in Fig.~\ref{fig:Results}.

On the right panel of Fig.~\ref{fig:Results}, we show the $95\%$ confidence level contours for $G_E^s$ and $G_M^s$ at $Q^2=0.1\rm\;GeV^2$ where multiple experimental values are available. As can be seen, we obtain a very precise bound on the values of the electric and magnetic form factors.

\section{Conclusions}
Our results on the strange electromagnetic form factors of
the nucleon are presented, obtained using four ensembles at four different lattice spacings of \Nf{2}{1}{1} twisted mass fermions and with physical pion mass. Plateau fits are performed to obtain the ground state matrix element where similar source-sink separation in physical units are used to obtain the final values of the form factors for each ensemble. In order to obtain a continuum extrapolation, we use a combined fit of the
$Q^2,a^2$-dependence using Galster-like, dipole and, $z$-expansion Ans\"atze. We obtain precise   results for the radii and magnetic moments. In Fig.~\ref{fig:Compare_lattice_results}, we present a comparison with different lattice results, where our results are the first ones in the continuum limit  using ensembles all at the physical point.  We additionally observe good agreement with the results obtained in Refs.~\cite{Leinweber:2004tc,Leinweber:2006ug} indirectly through a combination of experimental input and lattice QCD data for $G_E(Q^2 =0.1 \rm \;GeV^2)$, whereas $\mu^s$, while consistent in sign, is smaller in magnitude in this work.

\section*{Acknowledgments}
C.A., S.B., C.I., G.K., F.P., and G.S.  acknowledge partial support by the projects
Baryon8, MuonHVP, PulseQCD, DeNuTra, IMAGE-N, HyperON,  StrongILA and partonWF (POSTDOC/05\-24/0001, EXCELLENCE/0524/0017,
EXCELLENCE/0524/0269, EXCELLENCE/0524/0455, EX\-CELLENCE/0524/0459,
VISION ERC-PATH 2/0524/0001, EXCELLENCE/0524/0001 and VISION ERC/0525/0010, respectively)
co-financed by the European Regional Development Fund and the Republic
of Cyprus through the Research and Innovation Foundation as well as
AQTIVATE that received funding from the European Union’s research and
innovation program under the Marie Sklodowska-Curie Doctoral Networks
action, Grant Agreement No 101072344. C.A acknowledges support by the
University of Cyprus projects ``Nucleon-GPDs'' and
``PDFs-LQCD''. B.P. is supported by ENGAGE which received funding from
the EU's Horizon 2020 Research and Innovation Programme under the
Marie Skłodowska-Curie GA No. 101034267. This project received funding from the European Research Council (ERC) via the project "LEEX" grant agreement 101170304. Funded by the European Union. Views and opinions expressed are however those of the author(s) only and do not necessarily reflect those of the European Union or the European Research Council Executive Agency (ERCEA). Neither the European Union nor the ERCEA can be held responsible for them. This work was supported by
grants from the Swiss National Supercomputing Centre (CSCS) under
projects with ids s702 and s1174. The authors gratefully acknowledge
the Gauss Centre for Supercomputing e.V. (www.gauss-centre.eu) for
funding this project by providing computing time through the John von
Neumann Institute for Computing (NIC) on the GCS Supercomputer
JUWELS-Booster at J\"ulich Supercomputing Centre (JSC). This project
received access to the JUPITER supercomputer, which is funded by the
EuroHPC Joint Undertaking, the German Federal Ministry of Research,
Technology and Space, and the Ministry of Culture and Science of the
German state of North Rhine-Westphalia, through the JUPITER Research
and Early Access Program (JUREAP) as part of the EuroHPC project
application EHPC-EXT-2023E02-052. The authors also acknowledge the
Texas Advanced Computing Center (TACC) at University of Texas at
Austin for providing HPC resources.

\bibliographystyle{JHEP} \bibliography{skeleton}

\end{document}